\documentclass[12pt,preprint]{aastex631} 
\usepackage{color}
\usepackage{hhline}
\usepackage{graphicx}
\usepackage{booktabs}
\usepackage{threeparttable}
\usepackage{longtable}
\usepackage{rotating}
\usepackage{hyperref}
\usepackage{scalefnt}
\usepackage{amsmath}
\hypersetup{
    colorlinks=true,
    linkcolor=blue,
    filecolor=magenta,      
    urlcolor=blue,
}
\shorttitle{Fluorine Abundances in the Galactic Nuclear Star Cluster}
\shortauthors{Guer\c{c}o et al.}

\begin{document}

\title{Fluorine Abundances in the Galactic Nuclear Star Cluster}

\author[0000-0002-0151-5212]{Rafael Guer\c{c}o}
\affil{Observat\'orio Nacional, S\~ao Crist\'ov\~ao, Rio de Janeiro, Brazil}
\email{guercorafael@gmail.com}

\author{Solange Ram\'irez}
\affil{Carnegie Observatories, Pasadena, CA 91107, USA}

\author[0000-0001-6476-0576]{Katia Cunha}
\affil{University of Arizona, Tucson, AZ 85719, USA}
\affil{Observat\'orio Nacional, S\~ao Crist\'ov\~ao, Rio de Janeiro, Brazil}

\author[0000-0002-0134-2024]{Verne V. Smith}
\affil{NOIRLAB, Tucson, AZ 85719 USA}

\author[0000-0002-4591-6253]{Nikos Prantzos}
\affil{Institut d'Astrophysique de Paris, France}

\author[0000-0003-0817-2862]{Kris Sellgren}
\affil{The Ohio State University, Columbus, OH 43210, USA}

\author[0000-0001-9205-2307]{Simone Daflon}
\affil{Observat\'orio Nacional, S\~ao Crist\'ov\~ao, Rio de Janeiro, Brazil}

\accepted{March 08, 2022}
\submitjournal{The Astrophysical Journal}

\begin{abstract}
Abundances of fluorine ($^{19}$F), as well as isotopic ratios of $^{16}$O/$^{17}$O, are derived in a sample of luminous young ($\sim$10$^{7}$--10$^{8}$ yrs) red giants in the Galactic center (with galactocentric distances ranging from 0.6--30 pc), using high-resolution infrared spectra and vibration-rotation lines of H$^{19}$F near $\lambda$2.3$\mu$m.  Five of the six red giants are members of the Nuclear star cluster that orbits the central supermassive black hole.  Previous investigations of the chemical evolution of $^{19}$F in Galactic thin and thick disk stars have revealed that the nucleosynthetic origins of $^{19}$F may be rather complex, resulting from two, or more, astrophysical sites; fluorine abundances behave as a primary element with respect to Fe abundances for thick disk stars and as a secondary element in thin disk stars.  The Galactic center red giants analyzed fall within the thin disk relation of F with Fe, having near-solar, to slightly larger, abundances of Fe ($<$[Fe/H]$>$=+0.08$\pm$0.04), with a slight enhancement of the F/Fe abundance ratio ($<$[F/Fe]$>$=+0.28$\pm$0.17).  In terms of their F and Fe abundances, the Galactic center stars follow the thin disk population, which requires an efficient source of $^{19}$F that could be the winds from core-He burning Wolf Rayet stars, or thermally-pulsing AGB stars, or a combination of both.  The observed increase of [F/Fe] with increasing [Fe/H] found in thin disk and Galactic center stars is not predicted by any published chemical evolution models that are discussed, thus a quantitative understanding of yields from the various possible sources of $^{19}$F remains unknown.

\end{abstract}

\keywords{fluorine abundances, chemical abundances, red giants, galactic center}

\section{Introduction}
Mapping detailed chemical abundance patterns in the diverse stellar populations that compose the Milky Way can reveal details about the formation and evolution of both specific stellar populations, as well as that of the Milky Way itself, along with deeper insights into stellar nucleosynthesis and evolution.  Among the elements heavier than hydrogen and helium, the quartet of carbon (Z=6), nitrogen (Z=7), oxygen (Z=8), and neon (Z=10) are, by far, the most abundant in number (as well as mass) and are crucial elements in understanding chemical evolution across the universe.  Nestled within these four abundance peaks in the periodic table lies fluorine (Z=9), whose abundance is $\sim$four orders-of-magnitude smaller than its periodic table neighbors; from carbon and beyond, the solar system abundance of $^{19}$F (the sole stable fluorine isotope) remains the smallest until scandium (Z=21).  Despite its relatively low abundance, fluorine isotopes play a key role in important H-burning (proton capture) reactions involving the various CNO-cycles and Ne-Na cycles, which set the relative abundances of the isotopes of C, N, O, Ne, and Na.  Compared to its abundant neighbors, the astrophysical origins of $^{19}$F remain relatively unconstrained due to a combination of observational challenges, along with a variety of potential sites for its nucleosynthesis. 

The earliest abundance determinations of fluorine were in red giant stars (e.g., Jorissen, Smith \& Lambert \citeyear{jorissen1992}; Cunha et al. \citeyear{cunha2003}; Smith et al. \citeyear{smith2005}; Cunha, Smith \& Gibson \citeyear{cunha2008}; Yong et al. \citeyear{yong2008}) and highlighted the challenges in observing fluorine abundance indicators in stars, due to the combination of its low abundance and atomic structure; these early studies used vibration-rotation transitions of H$^{19}$F in the near-infrared (NIR) K-band, near $\lambda$2.3$\mu$m, which required high-resolution NIR spectrographs.  The early studies probed samples of bright field red giant branch (RGB) and cool asymptotic giant branch (AGB) stars, along with small numbers of luminous red giant members of globular clusters, the Galactic bulge, or the LMC.  As several studies derived moderate enhancements of fluorine abundances in TP-AGB stars (e.g., Jorissen, Smith \& Lambert \citeyear{jorissen1992}; Uttenthaler et al. \citeyear{uttenthaler2008}; Abia et al. \citeyear{abia2009}, \citeyear{abia2010}, \citeyear{abia2015}), these studies suggested that AGB stars were one possible source of $^{19}$F, but not necessarily the dominant source in the Galaxy (e.g., Spitoni et al. \citeyear{spitoni2018}; Grisoni et al. \citeyear{grisoni2020}).

In addition to observations of HF in cool red giant stars, the ground-state neutral fluorine line, F I at $\lambda$954\AA, was detected in the ISM along three lines-of-sight by Federman et al. \citeyearpar{federman2005} using the FUSE satellite.  Federman et al. \citeyearpar{federman2005} determined current ISM abundance ratios of F/Cl and F/O that were solar, within the measurement uncertainties. High-excitation ($\chi$=12.7-13.0 eV) F I optical lines were also detected in several extreme helium (EHe) stars by Pandey \citeyearpar{pandey2006}, and RCrB stars by Pandey et al. \citeyearpar{pandey2008}, who found F to be very overabundant, by $\sim$1000-8000 times, in these stars.  As EHe and RCrB stars are likely in the post-AGB phase of evolution, these results implicated the post-AGB phase of stellar evolution as a potential source of fluorine. Werner, Rauch \& Kruk \citeyearpar{wernerrauchkruk2006} and Werner \& Herwig \citeyearpar{wernerherwig2006} pushed fluorine detections to the extremely hot H-deficient post-AGB stars by analyzing the F VI line at $\lambda$1139.5\AA \ using FUSE, also finding large F overabundances. More recently, Bhowmick, Pandey \& Lambert \citeyearpar{bhowmick2020} determined fluorine abundances in a number of hot EHe stars using optical F II lines and again found very large overabundances which, when compared to the abundances of C, O, and Ne, suggest that $^{19}$F production in these stars results from the merger of a CO-white dwarf with a He-white dwarf (creating the hot EHe star), as modelled by Menon et al. (\citeyear{menon2013}; \citeyear{menon2019}), or Lauer et al. \citeyearpar{lauer2019}.

Although the observational studies noted above that derive fluorine overabundances in TP-AGB stars may point to them as net sources of $^{19}$F in the Galaxy (e.g., Jorissen, Smith \& Lambert \citeyear{jorissen1992}; Uttenthaler et al. \citeyear{uttenthaler2008}; Abia et al. \citeyear{abia2015}, \citeyear{abia2019}), as well as certain types of merging white dwarf binary systems (e.g., Clayton et al. \citeyear{clayton2007}; Bhowmick, Pandey \& Lambert \citeyear{bhowmick2020}; Menon et al. \citeyear{menon2013}; \citeyear{menon2019}; Lauer et al. \citeyear{lauer2019}), there remain additional viable sources of $^{19}$F that likely contribute to significant fluorine production in the Galaxy, yet remain difficult to quantify from observations. Such sources include the scattering of neutrinos off of $^{20}$Ne during SN II core collapse (neutrino nucleosynthesis, or the $\nu$-process), as discussed by Woosley \& Haxton \citeyearpar{woosley_haxton1988} and Woosley et al. \citeyearpar{woosley1990}. Massive stars undergoing large mass loss rates during core He-burning (Wolf Rayet stars) were suggested by Meynet and Arnould \citeyearpar{meynet_arnould2000} as $^{19}$F sources, although the efficacy of this source was studied in more detail by Palacios, Arnould \& Meynet \citeyearpar{palacios2005}. Modelling by Prantzos et al. \citeyearpar{prantzos2018} found that rapidly rotating high-mass stars can also synthesize fluorine-19 and may be a significant source in the Galaxy.  Several studies have included all of these potential sources of $^{19}$F and incorporated them into models of Galactic chemical evolution that predict trends of, typically, [F/Fe] versus [Fe/H], or [F/O] versus [O/H], e.g., Timmes et al. \citeyearpar{timmes1995}, Kobayashi et al. (\citeyear{kobayashi2011a}; \citeyear{kobayashi2011b}), Prantzos et al. \citeyearpar{prantzos2018}, Spitoni et al. \citeyearpar{spitoni2018}, or Grisoni et al. \citeyearpar{grisoni2020}.

Several recent observational studies have focused on determining the behavior of the fluorine abundance as a function of either the Fe or O abundances in order to constrain the chemical evolution of $^{19}$F in the Galaxy, e.g., J\"onsson et al. (\citeyear{jonsson2014a}; \citeyear{jonsson2014b}), Pilachowski and Pace \citeyearpar{Pilachowski_Pace2015}, J\"onsson et al. \citeyearpar{jonsson2017}, Guer\c{c}o et al. \citeyearpar{guerco2019}, or Ryde et al. \citeyearpar{ryde2020}.  All of these works observed red giant stars and used vibration-rotation lines of HF in the $\lambda$2.2-2.4$\mu$m, while J\"onsson et al. \citeyearpar{jonsson2014b} also included
pure rotational lines of HF near $\lambda$12$\mu$m.  A takeaway summary of the recent observational results would note that the behavior of fluorine abundances as a function of either the Fe or O abundances is not fixed and varies over metallicity and possibly stellar population, indicating more than one astrophysical source for $^{19}$F (e.g., Ryde et al. \citeyear{ryde2020}), as suggested by both the nuclear astrophysics and modelling.

Franco et al. \citeyearpar{franco2021} recently added a new approach to mapping F abundances by using ALMA to detect interstellar medium HF in absorption in a lensed star-forming galaxy at a redshift of z=4.42 to obtain a measurement of the F/H abundance in the young universe.
The detected line is the J=1--$>$0 transition of HF with a rest wavelength of $\lambda$243.2433$\mu$m.  
Values of [F/H] were estimated from ratios of N(HF)/N(H$_{2}$), with the authors noting that the derived abundance of F represents a lower limit.  The values of N(HF) from Franco et al. \citeyearpar{franco2021} translate to a lower limit of [F/H]$\sim$-1.4 in the ISM of this star-forming galaxy at z=4.42, or a time when the universe was roughly 1 Gyr old.  Franco et al. \citeyearpar{franco2021} use a chemical evolution model that includes the $\nu$-process, AGB stars, and rapidly-rotating massive WR stars, with the conclusion that in this young star-forming galaxy, rapidly-rotating massive WR stars are the major source of the observed $^{19}$F abundance.  In the adopted chemical evolution model from Franco et al. \citeyearpar{franco2021}, AGB stars and the $\nu$-process are important sources of fluorine, but their contributions occur at later times (corresponding to redshifts of z$\sim$2--3).   

Guer\c{c}o et al. \citeyearpar{guerco2019} provided a glimpse into the overall behavior of F abundances across the Milky Way via the analysis of a sample of red giants spanning both a range of Galactocentric distances (R$_{g}$), as well metallicities, and found different trends of [F/Fe] versus [Fe/H] for the geometric thin and thick disk red giants, as well as a hint that values of [F/Fe] might be elevated in the outer Galaxy.  These results highlight the need to probe $^{19}$F abundances in different stellar populations and locations in the Milky Way, such as the Galactic center.

The center of the Milky Way is host to a super-massive black hole, along with a Nuclear stellar cluster (NSC), which is a common configuration in most massive galaxies (e.g., Sch\"odel et al. \citeyear{schodel2014}). The NSC is massive (M$_{NSC}\sim$2.5x10$^{7}$M$_{\odot}$) and is composed of multiple stellar populations, with older red giant stars (having ages from 3 -- 8 Gyrs), along with young (10$^{7}$-10$^{8}$ yrs) luminous massive stars (e.g., Nogueras-Lara, Sch\"odel \& Neumayer \citeyear{nogueras-lara2021}). The primary goal of this study is to probe the fluorine abundances in a sample of red giants in the Galactic center (at distances from the center of $\sim$2.5 -- 30 pc) whose chemical abundances (not including $^{19}$F) have been the subject of previous studies (Carr, Sellgren \& Balachandran \citeyear{carr2000}; Ramirez et al. \citeyear{ramirez2000}; Cunha et al. \citeyear{cunha2007}; Najarro et al. \citeyear{najarro2009}; Davies et al. \citeyear{davies2009}; Rich et al. \citeyear{rich2017}; Do et al. \citeyear{do2018}; Thorsbro et al. \citeyear{thorsbro2020}; see also Bentley et al. \citeyear{bentley2022}) these stars sample a new and unique region of the Galaxy in which to map fluorine abundances.

\section{Observations}

The targets are members of star clusters residing in the center of Milky Way previously studied in Cunha et al. \citeyearpar{cunha2007} and have been the subject of other previous studies (Carr et al. \citeyear{carr2000}; Ramirez et al. \citeyear{ramirez2000}; Davies et al. \citeyear{davies2009}). Five of the stars belong to the Nuclear Cluster and one star is from the Quintuplet Cluster.
Table \ref{tab:parameters_center} summarizes the observations, gives the 2MASS IDs when available, and their J, H, and K magnitudes.

\begin{figure}[t!]
  \centering
  \includegraphics[width=1.00\textwidth]{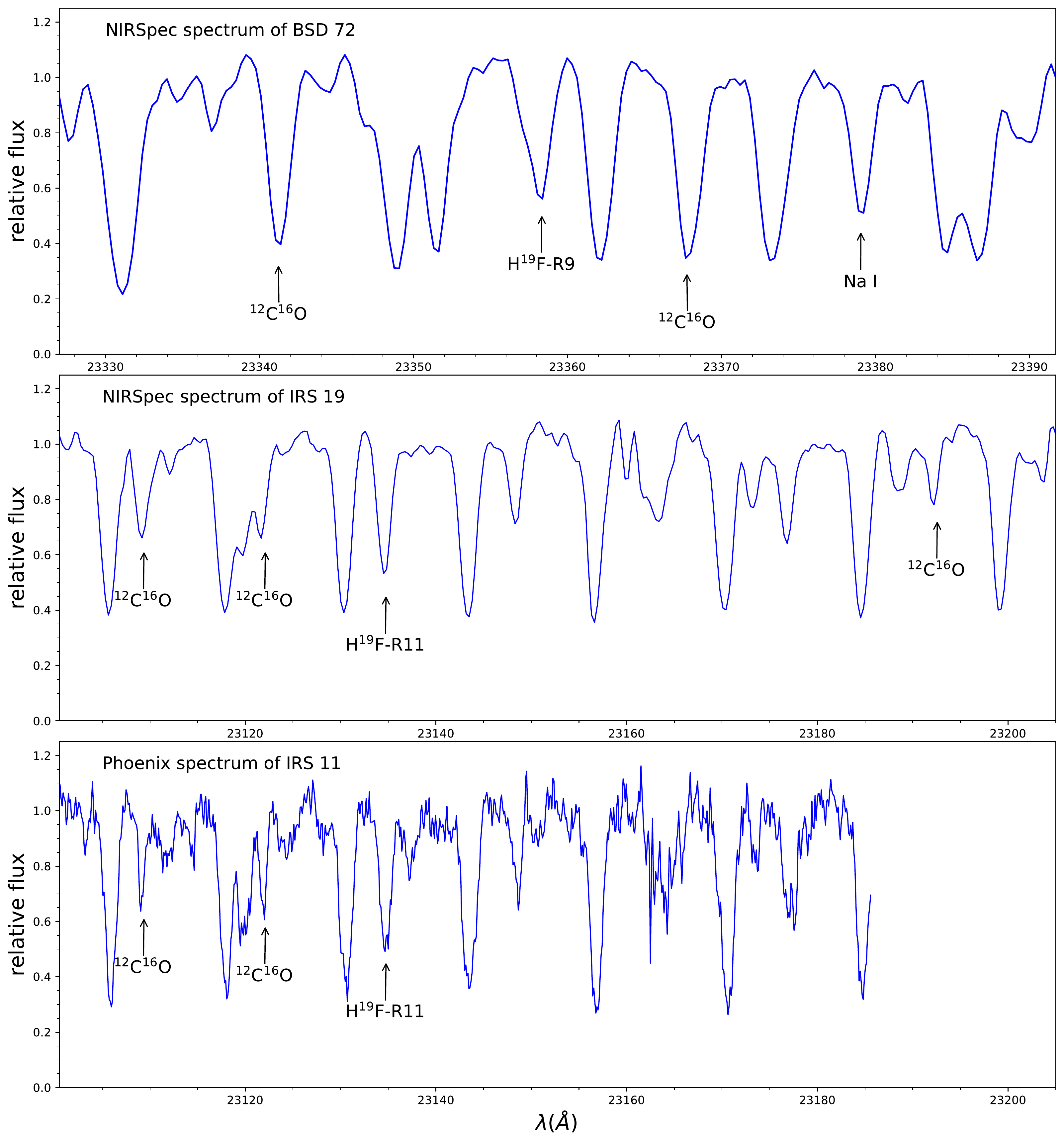}
  \caption{Sample spectra in the 2.3 $\mu$m region used in the abundance analysis. NIRSpec on the Keck II telescope spectra are shown in the top and middle panels and a Phoenix spectrum on the Gemini South telescope spectrum is shown in the bottom panel. The spectral region in the first panel includes the HF(1-0)R9 line, the Na I doublet ($\lambda$ 23378.945 and 23379.139 \AA) and two strong lines of CO ($\lambda$ 23341.22 and 23379.139 \AA); the spectral region in the middle and bottom panels includes the HF(1-0)R11 line and three weak lines of CO ($\lambda$ 23109.370, 23122.097 and 23192.581 \AA.)}
  \label{fig:spectra}
\end{figure}

The spectra analyzed here were obtained with the instruments NIRSpec (R=$\lambda$/$\Delta\lambda$=25,000) on the 10--meter Keck II telescope and Phoenix (R=$\lambda$/$\Delta\lambda$=50,000) on the 8--meter Gemini South telescope. 
The echelle order $\#$33 NIRSpec spectra, with lambda range between $\lambda$23,100\AA\ to $\lambda$23,420\AA were analyzed; the wavelength range for the Phoenix spectra are between $\lambda$23,095\AA\ to $\lambda$23,200\AA. 
The spectra were reduced in a standard way and followed the same basic steps and procedures as described, for example, in Cunha et al. \citeyearpar{cunha2007}.
In Figure \ref{fig:spectra} we show sample spectra of the Galactic sample targets and label the HF transitions along with CO lines analyzed (Table \ref{tab:lines}).
The top and middle panels show NIRSpec spectra including the HF-R9 and HF-R11 transitions, respectively. The bottom panel of Figure \ref{fig:spectra} shows a sample Phoenix spectrum in the H$^{19}$F-R11 region.

\section{Abundance Analysis} \label{sec:abundance}

The adopted values of effective temperature and surface gravity are presented in Table \ref{tab:parameters_center} and a corresponding Kiel diagram is presented in Figure \ref{fig:Kiel}; the Galactic center targets are cool (T$_{\rm eff}$ ranging from 3600K -- 3850K), evolved, and quite luminous, having log g values less than $\sim$0.4, with the exception of one target star that is less luminous (with log g=0.8).
The stellar parameters adopted for the studied targets were taken from Cunha et al. \citeyearpar{cunha2007}. The effective temperatures in that study were based on spectral indices of CO and H$_2$O absorption, with the final T$_{\rm eff}$ being an average of results derived originally by Blum et al.\citeyearpar{blum2003} and those obtained from a relation they constructed of T$_{\rm eff}$ with spectral type, which was based on effective temperatures derived in Smith \& Lambert \citeyearpar{smith_lambert1990}.
Log g values were obtained using the absolute bolometric magnitudes from Blum et al. \citeyearpar{blum2003}.

Fluorine abundances (Table \ref{tab:abundances}) were derived by calculating synthetic spectra in local thermal equilibrium (LTE) with the code Turbospectrum (TS; Alvarez \& Plez \citeyear{alvarez1998}; Plez \citeyear{plez2012}) using the MARCS models atmosphere (Gustafsson et al. \citeyear{gustafsson2008}) that we interpolate using the online grid of OSMARCS models (\url{http://marcs.astro.uu.se/}) corresponding to  5 - 15 M$_\odot$. 
To compute best fits to the observed spectra, the BACCHUS wrapper (Masseron, Merle \& Hawkins \citeyear{masseron2016}) was used in manual mode, including a single Gaussian to represent both the instrumental broadening and the macroturbulent velocity. 

The derivations of the fluorine abundances were based on two vibrational–rotational HF (1--0) R9 and R11 transitions at $\lambda$2.3$\mu$m. The HF lines used in the abundance analyses are presented in the Table \ref{tab:lines}, with their excitation potentials ($\chi$; from J\"onsson et al. \citeyear{jonsson2014a} and Decin \citeyear{decin2000}), log gf-values (from J\"onsson et al. \citeyear{jonsson2014a}), and the dissociation energy (D$_\circ$ = 5.869 eV; from Sauval \& Tatum \citeyear{sauval_tatum1984}).  Five strong CO lines found in the same spectral region where the HF lines reside (Table \ref{tab:lines}) were used in order to estimate the microturbulent velocity parameter (see discussion below) adjustments were made to the carbon abundances as needed in order to obtain good fits to the CO lines. 

Abundance analyses based upon static model atmospheres typically require an additional small-scale line-broadening mechanism, usually added in as an extra velocity-broadening term referred to as microturbulence.  The value of the microturbulent velocity ($\xi$) in a particular star is set by the requirement that the derived abundance from a set of spectral lines of a specific species yield abundances that are independent of the line strength (more properly, the reduced equivalent width, W$_{\lambda}$/$\lambda$).  Although usually assumed to be modelled by a single value, it has been found in previous analyses of cool luminous red giants that low-excitation, strong molecular lines, such as from CO or OH, require larger values of the microturbulent velocity when compared to higher excitation (and typically weaker) atomic or molecular lines (e.g., Smith \& Lambert \citeyear{smith_lambert1986}, \citeyear{smith_lambert1990}; Tsuji \citeyear{tsuji1988}, \citeyear{tsuji1991}, \citeyear{tsuji2008}).  As the HF lines analyzed in the Galactic center stars are low excitation and strong, values for $\xi$ were set using a set of $^{12}$C$^{16}$O lines that have a range of line strengths which overlap those of the H$^{19}$F lines.  The microturbulent velocities are included in Table \ref{tab:parameters_center} and range from $\xi$= 2.3 to 4.8 km$\cdot$s$^{-1}$, similar to values derived by Carr et al. \citeyearpar{carr2000} for the Galactic center luminous cool supergiant IRS7 ($\xi$=3.0 km$\cdot$s$^{-1}$), and the supergiants $\alpha$ Ori ($\xi$=3.1--3.8 km$\cdot$s$^{-1}$) and VV Cep ($\xi$=3.5--4.4 km$\cdot$s$^{-1}$).  

In addition to the numerous and strong $^{12}$C$^{16}$O lines falling within the spectral region of the HF lines, there are also lines from $^{12}$C$^{17}$O(2--0) running through the spectra around the HF R9 line, with this line itself blended with the $^{12}$C$^{17}$O(2--0) R25 line.  Dredge-up on the red giant branch will bring an increased abundance of $^{17}$O to the surface (along with $^{13}$C), with ratios of $^{16}$O/$^{17}$O decreasing from solar-like ratios of $\sim$2750, to values as low as $\sim$400-100 for masses greater than 2M$_{\odot}$ (for recent studies of $^{16}$O/$^{17}$O in various types of red giants, see Lebzelter et al. \citeyear{lebzelter2019a}, \citeyear{lebzelter2019b}).  Since this sample of luminous red giants consists of rather massive stars (M$>$5M$_{\odot}$), dredge-up on the red giant branch will result in detectable $^{12}$C$^{17}$O lines near the H$^{19}$F R9 line.  In order to improve the fit to the H$^{19}$F R9 line, it was necessary to determine the abundance of $^{17}$O via the CO lines and these abundances, as ratios of $^{16}$O/$^{17}$O, are presented in Table \ref{tab:abundances}.  These ratios were derived from combinations of the $^{12}$C$^{17}$O(2--0)R25, R26, R27, R28, or R29 lines.

\subsection{Abundance Uncertainties}

Uncertainties in the derived F and C abundances were estimated establishing the sensitivity of these abundances to changes in the stellar parameters corresponding to their typical errors in T$_{\rm eff}$ ($\pm$100 K), $\log g$ ($\pm$0.25 dex), metallicity ($\pm$0.10 dex), and microturbulent velocity ($\pm$0.25 km$\cdot$s$^{-1}$). The estimated uncertainties for each parameter and the square root of the sum in quadrature of the corresponding abundance ($\Delta$A $= [(\delta T)^2 + (\delta \log g)^2 + (\delta [Fe/H])^2 + (\delta \xi)^2]^{1/2}$) are presented in the Table \ref{tab:disturbance}. We used a baseline model atmosphere with T$\rm _{eff}$ = 3700 K, log g = 0.50, [Fe/H]=0.00, and $\xi$ = 4.50 km/s to calculate the sensitivities of abundances with the disturbance $\delta T$, $\delta \log g$, $\delta [Fe/H]$ and $\delta \xi$.

\section{Discussion}

\begin{figure}[t!]
  \centering
  \includegraphics[width=1.0\textwidth]{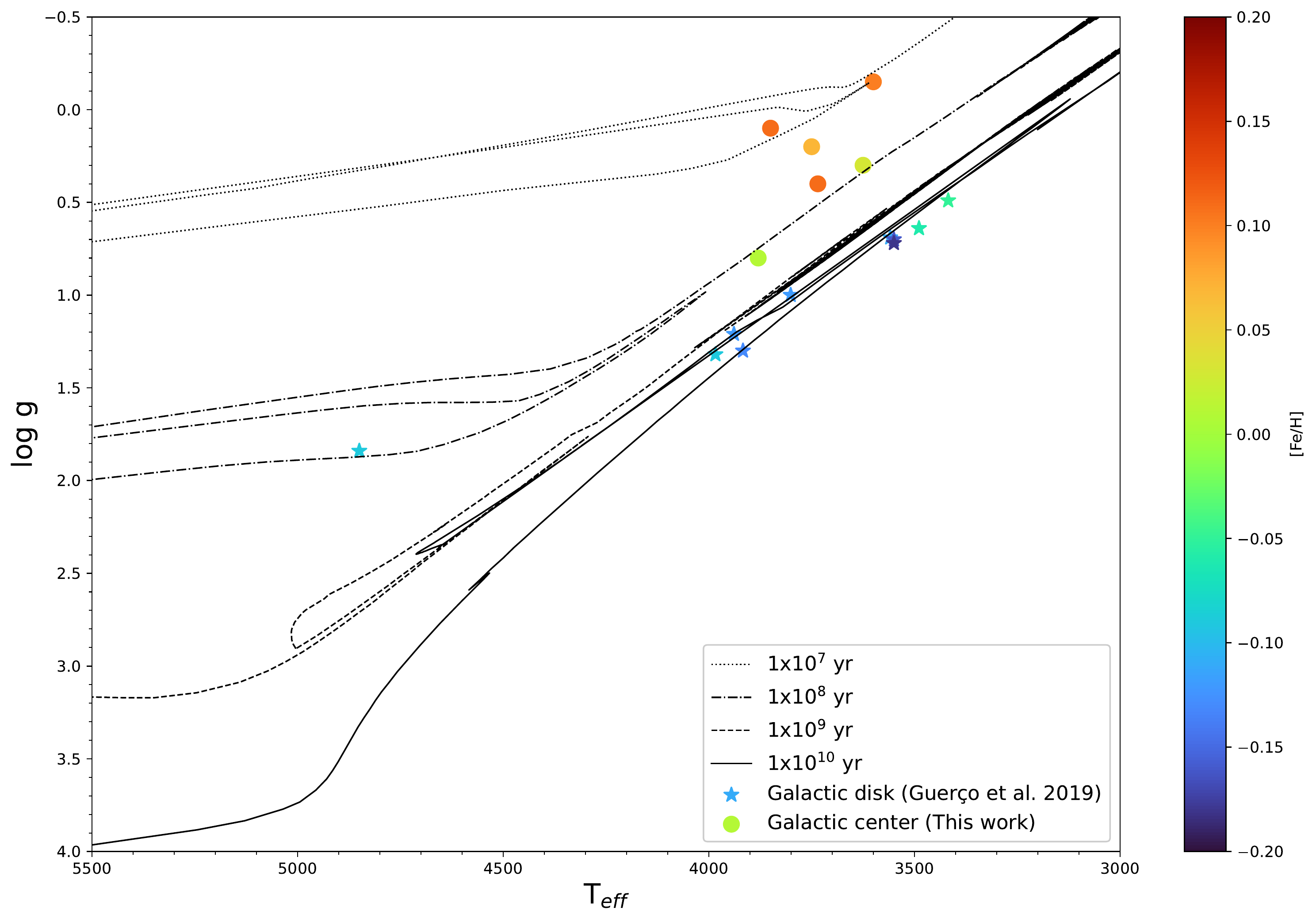}
  \caption{T$\rm_{eff}$-$\log g$ diagram (Kiel diagram) for the luminous red giants in this sample are shown as filled circles; asterics represent a sub-sample of stars from Guerço et al. \citeyearpar{guerco2019} having metallicities around solar; isochrones  (from Bressan et al. \citeyear{bressan2012}) for solar metallicities and ages 1$\times 10^7$ yr (dotted line), 1$\times 10^8$ yr (dot-dashed line), 1$\times 10^9$ (dashed line) yr and 1$\times 10^{10}$ yr (solid line) are also shown to illustrate relative positions on the RGB. The stars are color-coded by metallicity between -0.2 and 0.2 dex.}
  \label{fig:Kiel}
\end{figure}

Figure \ref{fig:Kiel} shows the Kiel diagram with the stellar parameters adopted in this study for the sample of Galactic center stars (filled circles).
We also show for comparison those stars with metallicities $>$ -0.2 dex from the disk sample analyzed previously in Guer\c{c}o et al. \citeyearpar{guerco2019}, as these will be used to define the disk trend around solar metallicities in the discussion of fluorine results. 
It is apparent from a comparison with the solar metallicity isochrones shown (Bressan et al. \citeyear{bressan2012}) that the Galactic center targets are much younger and, thus, more massive, having ages between 10$^{7}$ and 10$^{8}$ years, and masses in the range of M$\sim$6-16M$_{\odot}$ (Cunha et al. \citeyear{cunha2007}), when compared to the stars from the Galactic disk previously studied, which span ages from 10$^{9}$ - 10$^{10}$ years and masses of M$\sim$1-2M$_{\odot}$. The more metal poor disk targets in Guer\c{c}o et al. \citeyearpar{guerco2019} are also older than the Galactic center population, as can be seen from Figure \ref{fig:Kiel} in that paper.  

\subsection{Fluorine in the Galactic Center and the Disk}

\begin{figure}[t!]
  \centering
  \includegraphics[width=1.0\textwidth]{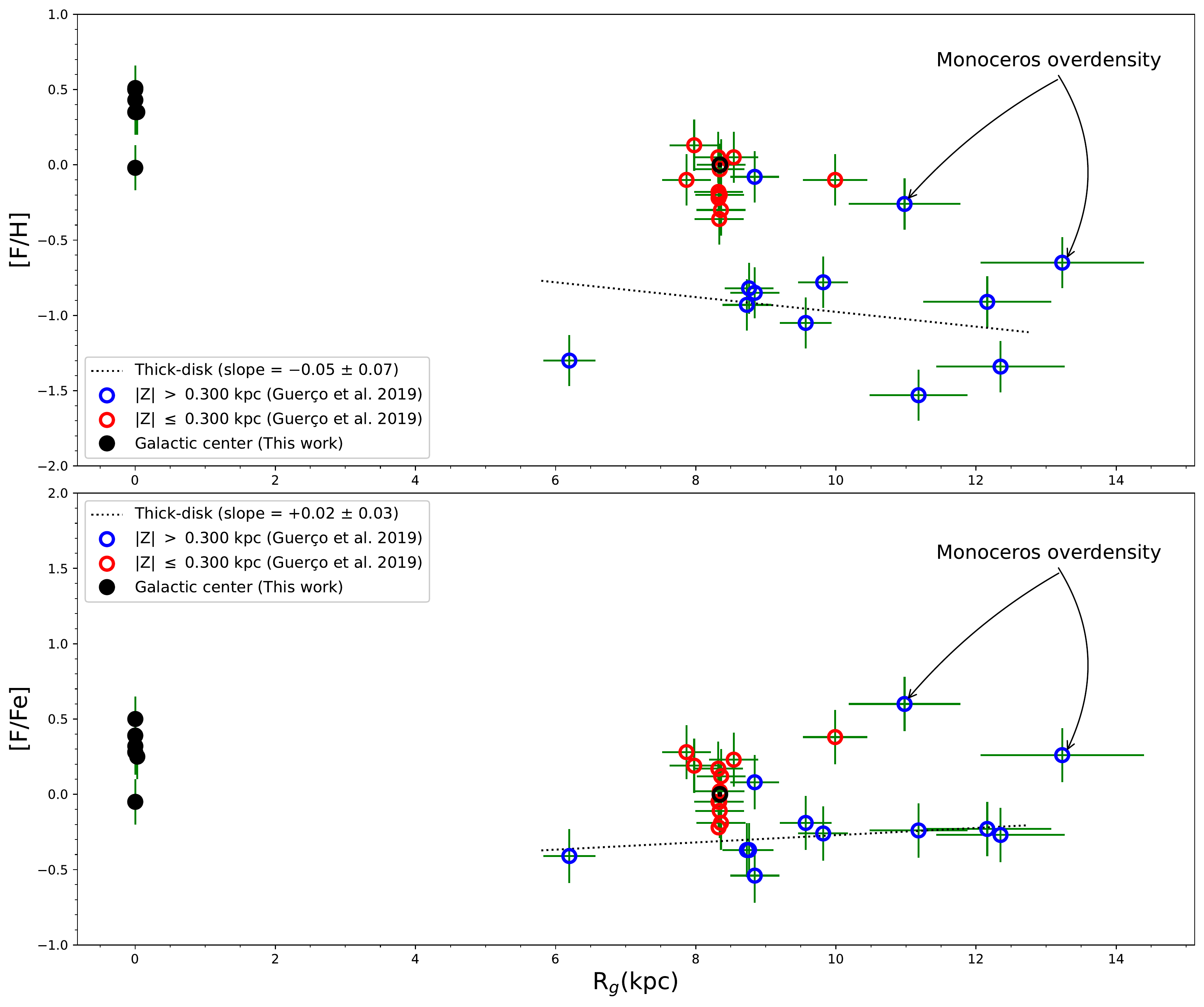}
  \caption{Top panel: [F/H] abundances as a function of the projected galactocentric distance, Rg; bottom panel: [F/Fe] abundances as a function of Rg. Black circles represent the Galactic center stars analyzed. The open circles are disk stars from Guerço et al \citeyearpar{guerco2019}.
  The dotted line represents the best-fit slope to disk stars with distances to the mid plane $>$ 300 pc (blue circles taken as thick-disk stars, except the Monoceros overdensity stars). The distances of the Galactic center stars are from Cunha et al. \citeyearpar{cunha2007}. The distances and uncertainties of the disk stars are based on Gaia EDR3 (Gaia Collaboration et al. \citeyear{gaia2021}) and were taken from Bailer-Jones et al. \citeyearpar{bailer-jones2021}. The solar references for fluorine and iron are from Maiorca et al. \citeyearpar{maiorca2014} and Asplund et al. \citeyearpar{asplund2009}.}
  \label{fig:gradient}
\end{figure}

The [F/H] abundances obtained here for the sample of stars residing in the Galactic center, in conjunction with previous results obtained for disk stars, provide the opportunity to begin to explore possible radial variations in fluorine abundances in the Galactic disk. 
Within this context, it is important to note that rotating and non-rotating massive star models predict that fluorine abundances in stars having masses within the studied range are not expected to be destroyed (Ekstr\"om et al. \citeyear{ekstrom2012}) and therefore their fluorine content would represent their natal cloud value.
Figure \ref{fig:gradient} shows the [F/H] (top panel) and [F/Fe] (bottom panel) abundances versus galactocentric distances for fluorine results obtained homogeneously here (filled black circles) and in Guer\c{c}o et al. \citeyearpar{guerco2019}; the red open circles represent stars which are probable thin disk members and the blue open circles probably belong to the thick disk, noting that the distinction between thick and thin in Guer\c{c}o et al. \citeyearpar{guerco2019} comes solely from the distance, Z, a star has from the Galactic mid-plane. 
Given the uncertainties in the fluorine abundance determinations (see sensitivity of the HF abundances to stellar parameters in Table \ref{tab:disturbance}), the fluorine content around the solar neighborhood (for stars within $\pm$ 1 kpc from the Sun) is roughly solar: $<$[F/H]$>$ = -0.09 $\pm$ 0.15 and $<$[F/Fe]$>$ = 0.09 $\pm$ 0.25. (We use the solar fluorine of A(F)= 4.40 $\pm$ 0.25 from Maiorca et al. \citeyear{maiorca2014} in this comparison.) The results for Galactic center population are slightly enhanced $<$[F/H]$>$ = 0.35 $\pm$ 0.18 and $<$[F/Fe]$>$ = 0.28 $\pm$ 0.17.

Guer\c{c}o et al. \citeyearpar{guerco2019} trace a fluorine gradient for their probable thick disk stars that have galactocentric distances between roughly 6 and 12 kpc; the best fit slope to their data is also shown in Figure \ref{fig:gradient}. At this point, however, it is not possible to reliably trace a fluorine gradient for the thin disk, first because the fluorine results at hand do not extend towards the inner Galaxy and, second, because abundance gradients are expected to and have been shown to evolve with time (e.g., Minchev, Chiappini \& Martig \citeyear{minchev2013}, Donor et al. \citeyear{donor2020}, Spina et al. \citeyear{spina2021}), thus we would be mixing fluorine results for a young population with those of field disk stars of unknown ages.  Simply combining the fluorine abundances for the thin disk, however, would result in a nearly null variation with galactocentric distance for the [F/Fe] ratio and a small negative slope for [F/H]. However, as discussed in Cunha et al. \citeyearpar{cunha2007} for iron, for example, the extrapolation of the metallicity gradient for the inner galaxy would predict a much higher metallicity for the Galactic center, which is not observed; the metallicity of the Galactic center young population is roughly solar or slightly above (Ramirez et al. \citeyear{ramirez2000}, Davies et al. \citeyear{davies2009}, etc).

\subsection{Chemical Evolution of Fluorine}

\begin{figure}[t!]
  \centering
  \includegraphics[width=1.0\textwidth]{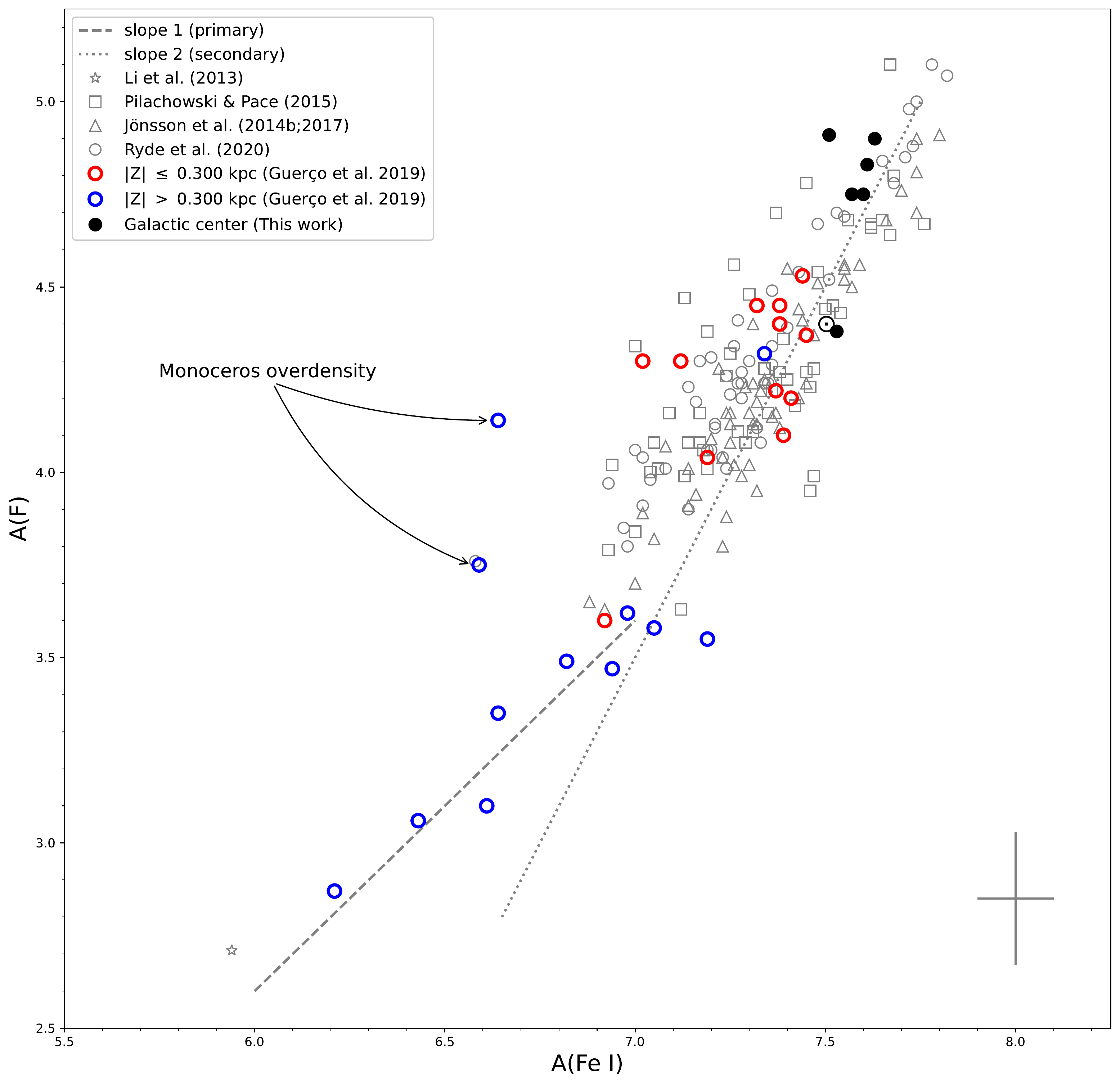}
  \caption{Behavior of the fluorine abundance with metallicity for the studied stars and previous results from  the literature. Dashed lines representing pure primary and secondary behaviors for the change of fluorine with metallicity are also shown. The stars of Guer\c{c}o et al. \citeyearpar{guerco2019} are segregated according to their distances, Z, from the Galactic mid-plane; the open blue circles correspond to probable thick-disk/halo stars and the open red circles to probable thin-disk stars. Two stars identified as possible members of the Monoceros overdensity in Guer\c{c}o et al. \citeyearpar{guerco2019} are marked. The black circles correspond to the results obtained in this work for the Galactic center stars. A representative error bar for the abundances is shown. The solar references for fluorine and iron are from Maiorca et al. \citeyearpar{maiorca2014} and Asplund et al. \citeyearpar{asplund2009}.}
  \label{fig:Fe_vs_F}
\end{figure}

Fluorine abundances as a function of metallicity from main studies in the literature are summarized in Figure \ref{fig:Fe_vs_F}. Again the black filled circles represent the Galactic center results from this study, while the open circles are the disk sample from Guer\c{c}o et al. \citeyearpar{guerco2019}, discussed above. We use the same symbols as before: in red are shown those stars having distances from the mid plane Z $<$ 300 pc and in blue those with Z $>$ 300 pc.

Other results from the literature are also shown in Figure \ref{fig:Fe_vs_F} as grey open symbols (Li et al. \citeyear{li2013}; J\"onsson et al. \citeyear{jonsson2014b}, \citeyear{jonsson2017}; Pilachowski \& Pace \citeyear{Pilachowski_Pace2015}; Ryde et al. \citeyear{ryde2020}). The overlap between the derived metallicities for A(Fe)$>$7.0 ([Fe/H]$>$-0.5), where most of the results from the literature are found, is clear. 
A key aspect of the F versus Fe abundances in Figure \ref{fig:Fe_vs_F}, as noted and emphasized by Guer\c{c}o et al. \citeyearpar{guerco2019}, is the distinctive behaviors of A(F) as a function of A(Fe) at lower metallicities when compared to higher metallicities (with A(Fe)$\sim$6.9-7.0 being the ``transition'' metallicity). The two lines in Figure \ref{fig:Fe_vs_F} illustrate slopes of 1 (dashed) and 2 (dotted), which delineate primary and secondary yields, respectively, of $^{19}$F when compared to Fe; the fluorine observed in the lower metallicity stars arises from a primary-like source, while fluorine in the higher metallicity stars is dominated by a secondary-like source.
In particular, the Galactic center fluorine abundances fall roughly within the distribution of results for those stars with metallicities above solar in the samples of Pilachowski \& Pace \citeyearpar{Pilachowski_Pace2015}, J\"onsson et al. (\citeyear{jonsson2014b}, \citeyear{jonsson2017}), or Ryde et al. \citeyearpar{ryde2020}. 
Since the Galactic center stars sampled here are quite young, span a small range in metallicity, and may represent a distinct population, it is not possible to conclude that the same secondary-like source (or sources) that drives the chemical evolution in the thin disk is at work, but simply note that the Galactic center fluorine abundances fall within the distribution of secondary-like behavior in disk stars.

There are uncertainties in both the derived F and Fe abundances and the impact of these uncertainties should be included as part of this discussion.  As noted above, the young Galactic center stars, based upon the analysis here, are slightly more metal-rich than solar, with $<[Fe/H]>\sim$+0.08 and $<[F/H]>\sim$+0.35. Since the F abundances are much more sensitive to T$_{\rm eff}$ than the Fe abundances, a small offset of -50K in T$_{\rm eff}$ would result in values of both [Fe/H] and [F/H]$\sim$+0.1, with [F/Fe]$\sim$0.0.  Our results, when viewed in light of likely uncertainties in combinations of T$_{\rm eff}$ and log g, indicate that the young Galactic center stars have near-solar Fe and F abundances to slightly metal-rich stars with modest enhancements of F, with the stars falling along the secondary-like trend of fluorine relative to Fe.

\begin{figure}[t!]
  \centering
  \includegraphics[width=1.0\textwidth]{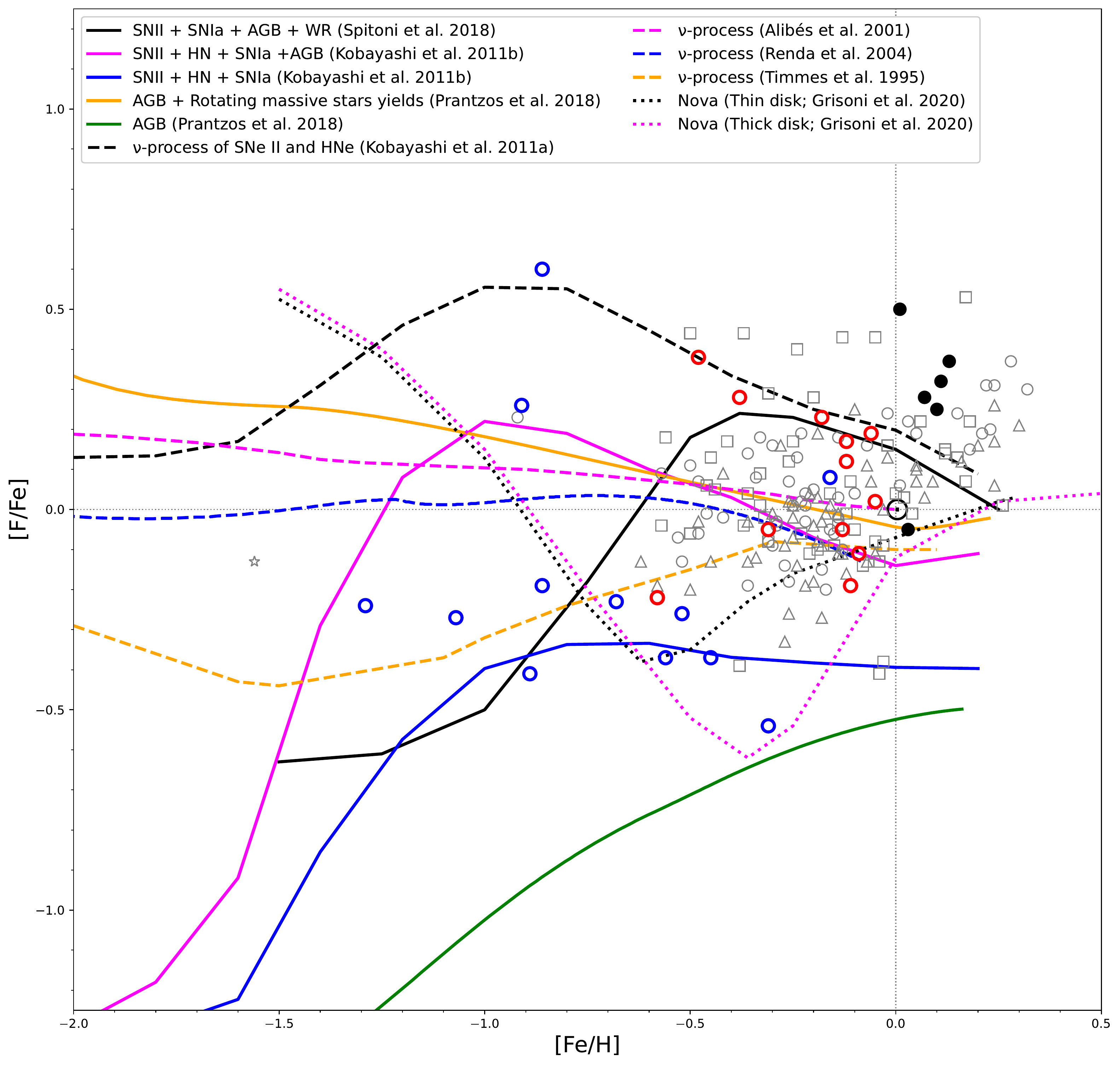}
  \caption{Chemical evolution of fluorine viewed as [F/Fe] vs. [Fe/H].  The black circles correspond to the Galactic center stars from this work; the open blue circles correspond to probable thick-disk/halo stars and the open red circles to probable thin-disk stars from Guer\c{c}o et al. \citeyearpar{guerco2019}; all computed in the same scale as in this study.  Results from other studies in the literature are shown as grey symbols (same as for Figure \ref{fig:Fe_vs_F}). Several chemical evolution models from the literature are also shown.  The solar references for fluorine and iron are from Maiorca et al. \citeyearpar{maiorca2014} and Asplund et al. \citeyearpar{asplund2009}.}
  \label{fig:met_vs_FFe}
\end{figure}

Figure \ref{fig:met_vs_FFe} illustrates a variety of Galactic chemical evolution models that include various sources for the nucleosynthesis of $^{19}$F, plotting [F/Fe] versus [Fe/H], and including the abundances derived here for the Galactic center stars shown as filled black circles. Also included are the abundances from Guer\c{c}o et al. \citeyearpar{guerco2019} and the other observational studies, as discussed in the previous paragraph, the various observational studies are in good agreement.  
Guer\c{c}o et al. \citeyearpar{guerco2019} discussed most of the chemical evolution models shown in Figure \ref{fig:met_vs_FFe} in some detail, and here we review briefly these models and where the Galactic center abundances fit in. With [F/Fe] plotted as a function of [Fe/H] in Figure \ref{fig:met_vs_FFe}, primary chemical evolution would be flat (slope$\sim$0), while secondary would have a slope of 1. In general, this is the behavior seen in the derived abundance ratios from the sample in Guer\c{c}o et al. \citeyearpar{guerco2019}. 
Alas, as discussed previously, the Galactic center young stars, given their small range in metallicity, do not shed new light on the primary-like and secondary-like $^{19}$F sources other than the fact that their [F/Fe] values are not in contrast with those of disk stars at the same iron abundances.

The rather large number of chemical evolution models displayed in Figure \ref{fig:met_vs_FFe} may look bewildering at first glance, but do illustrate the significant uncertainties regarding the origin(s) of fluorine and the corresponding stellar yields,
and careful inspection can simplify the behaviors of the various models. 
(See also the discussion in Grisoni et al. \citeyear{grisoni2020}).
In general, most of the models in Figure \ref{fig:met_vs_FFe} tend to converge to having roughly solar [F/Fe] (with boundaries at roughly [F/Fe]= +0.2 and -0.1) at [Fe/H]$\sim$0.0. This is not unexpected as the solar reference is commonly used to constrain models of the solar neighborhood
and, in fact, the fluorine-over-iron abundances for solar neighborhood stars also scatter around solar (as also shown in Figure \ref{fig:Fe_vs_F}). 
In addition to the models shown in Figure 5, Olive \& Vangioni (2019) scale neutrino process contributions using constraints from observed B11/B10 ratios, and find that AGB stars are the main contributors at solar metallicity, obtaining a flat and roughly solar behavior for [F/Fe] between [Fe/H]$\sim$-1 and solar.

Focusing solely on the predicted model values of [F/Fe] for metallicities above solar, we note that all the above 1-zone models were constructed mainly for the solar neighborhood; stars with super-solar metallicities now found in the solar neighborhood, are thought to have originated from the inner disk, which has super-solar metallicity gas, via radial migration and in a time-scale of a few Gyr. The models shown therefore may not be as meaningful in such high-metallicity regimes.
The models by Kobayashi et al. \citeyearpar{kobayashi2011b} and Spitoni et al. \citeyearpar{spitoni2018} are the only ones that predict enhanced values of [F/Fe] at metallicities slightly above solar, as seen for the Galactic center stars, but such models find a decrease in [F/Fe] as the metallicity increases. 
All of the other models shown in Figure \ref{fig:met_vs_FFe} do not reach values of [F/Fe] above solar for [Fe/H] $>$ 0. 
The fluorine results shown in Figure \ref{fig:met_vs_FFe} corroborate this point: stars in the different fluorine studies (Ryde et al. \citeyear{ryde2020}; J\"onsson et al. \citeyear{jonsson2017}; Pilachowski \& Pace \citeyear{Pilachowski_Pace2015}) having metallicities of solar and above (0.0 $<$ [Fe/H] $<$ +0.3) also show increased values of [F/Fe], an increase in [F/Fe] with metallicity for Fe/H $>$ 0  as seen in the fluorine data for disk stars and the Galactic center is not predicted by any of the models.

Focusing on the predictions of [F/Fe] at the lowest metallicities shown in Figure \ref{fig:met_vs_FFe} ([Fe/H]$<$-1.7) does show that all of the models predict primary-like evolution in [F/Fe]. In fact, the solid black (Spitoni et al. \citeyear{spitoni2018}), magenta and blue (Kobayashi et al. \citeyear{kobayashi2011a}) lines show secondary behaviour; Timmes et al. \citeyearpar{timmes1995}, Alib\'es, Labay \& Canal \citeyearpar{alibes2001}, Renda et al. \citeyearpar{renda2004}, and Kobayashi et al. \citeyearpar{kobayashi2011b} (see also Olive \& Vangioni \citeyear{Olive2019}) find the $\nu$-process to be a source of primary $^{19}$F at these low metallicities. 

Prantzos et al. \citeyearpar{prantzos2018} adopt yields from rapidly rotating massive stars from Limongi \& Chieffi \citeyearpar{limongi2018}, where rotationally-induced mixing results in the synthesis of primary $^{14}$N and subsequently primary $^{19}$F in He-burning. This is an important difference between the neutrino-induced nucleosynthesis and rotationally induced mixing;  the former produces primary F from neutrino spallation of Ne, but does not affect N which is normally made as secondary from C+O; in contrast, the latter produces both N and F as primaries, and this is a distinctive signature of rotation induced nucleosynthesis.

Given these models, with suitable adjustments to their yields and/or prescriptions for chemical evolution model details, a primary-like behavior of [F/Fe] in the lower-metallicity thick disk population would result from either the $\nu$-process or rapidly-rotating massive stars, or a combination of the two.
But the observed primary-like behaviour of N in the low-metallicity Galaxy (halo and thick disk) convincingly suggests that rotation is at play at low metallicities and provides an economic way to explain both N and F, if the latter really behaves as primary. The $\nu$-mechanism is more uncertain, basically because of the uncertainty in the neutrino spectra of CCSN.
The two stars at low metallicity with rather high values of [F/Fe] are ignored in this discussion, as they have been reported to be possible members of the Monoceros overdensity, which has been identified as a metal-poor stellar overdensity found $\sim$1-3 kpc out of the Galactic plane (Crane et al. \citeyear{crane2003}; Ibata et al. \citeyear{ibata2003}; Rocha-Pinto et al. \citeyear{rochapinto2003}; Morganson et al. \citeyear{morganson2016}). If members of an independent stellar population, these two stars likely formed from gas which has undergone chemical evolution that is distinct from either of the thick or thin disk populations.  

On the other hand, high rotation cannot be invoked at slightly sub-solar and $\sim$solar metallicities, because s-elements are highly overproduced (see discussion in Prantzos et al. \citeyear{prantzos2018}). AGB stars or, perhaps, novae could be invoked for those metallicities. However, it should be noticed that detailed massive star yields for supersolar initial metallicities do not exist, with the exception of Nomoto, Kobayashi \& Tominaga \citeyearpar{monoto2013}.
As mentioned above, the secondary-like evolution of [F/Fe] in the more metal-rich thin disk stars, including the mono metallicity young stars in the Galactic center studied here, is not accounted for by any of the models, although the increasing  values of [F/Fe] in the late thin disk and the Galactic center  require some other  sources, TP-AGB stars, novae and/or WR stellar winds.

One speculative source that is not included in the models noted here would be the H-deficient extreme He stars found by Pandey \citeyearpar{pandey2006}, Werner \& Herwig \citeyearpar{wernerherwig2006}, Pandey, Lambert \& Kameswara Rao \citeyearpar{pandey2008}, or Bhowmick, Pandey \& Lambert \citeyearpar{bhowmick2020} to be enormously overabundant in fluorine, which probably result from double-degenerate mergers between a CO white dwarf and a He white dwarf (Menon et al. \citeyear{menon2013}, \citeyear{menon2019}; Lauer et al. \citeyear{lauer2019}). Such systems may be too rare to have an impact on the Galactic $^{19}$F budget, but might be worth investigating.

\section{Conclusions}

Fluorine abundances have been derived in six massive young luminous red giants in the Galactic center; five of the stars are members of the Nuclear Star Cluster (NSC), which is centered on the central super-massive black hole (SMBH), while the sixth is a member of the Quintuplet Cluster (offset from the SMBH by $\sim$30 pc).  The NSC is massive ($\sim$2.5x10$^{7}$M$_{\odot}$), containing several generations of stars with ages ranging from $\sim$10 Gyr to $<$10$^{7}$ yrs (Nogueras-Lara, Sch\"odel \& Neumayer \citeyear{nogueras-lara2021}), and a significant spread in metallicity (Thorsbro et al. \citeyear{thorsbro2020}).  The luminous red giants analyzed here are young and evolved, with ages $\sim$10$^{7}$ to 10$^{8}$ yrs and masses of M$\sim$ 6-16M$_{\odot}$. The $^{19}$F abundances presented in this study are the first to be measured in the Galactic center, providing a view into the chemical evolution of fluorine in a unique region and population of the Milky Way.  A by-product of the analysis of the spectral region near the HF(1-0) R9 line is the derivation of $^{16}$O/$^{17}$O ratios; the RGB dredge-up of $^{17}$O is a function of stellar mass making the $^{16}$O/$^{17}$O ratios a sensitive probe of stellar nucleosynthesis and mixing along the RGB (e.g., Lebzelter et al. \citeyear{lebzelter2015}). A more extensive study of the CNO isotopic abundances of the Galactic center red giants is in preparation.

The population of Galactic center luminous young red giants have been analyzed for their Fe abundances in previous studies (Ramirez et al. \citeyear{ramirez2000}; Cunha et al. \citeyear{cunha2007}; Davies et al. \citeyear{davies2009};, Najarro et al. \citeyear{najarro2009}), all of which have found near-solar to slightly elevated [Fe/H] abundances. In particular, Cunha et al. \citeyearpar{cunha2007} derived iron abundances for the cluster members studied here and, when combined with our F abundances, results in a mean value of  $<$[F/Fe]$>$=+0.28$\pm$0.17;
these [F/Fe] ratios are roughly within the scatter of values for those stars previously studied near the Sun (with R$_{gal}\sim$8--10 kpc). The small evolution in the [F/Fe] ratios between the solar neighborhood and the Galactic center suggest that any radial gradients in [F/Fe] are likely small in the inner Galaxy, although it should be kept in mind that field stars in the inner galactic region between the solar neighborhood and the center of the Milky Way have yet to be probed.

Concerning the behavior of fluorine versus metallicity, we find that the fluorine abundances in the Galactic center sample generally overlap with those of disk stars at similar metallicities, with the latter following a secondary-like trend compared to Fe
(e.g., Guer\c{c}o et al. \citeyear{guerco2019}; Ryde et al. \citeyear{ryde2020}); [F/Fe] ratios are found to be marginally elevated in this sample of Galactic center stars, pointing to increasing [F/Fe] ratios with increasing [Fe/H].
Since the Galactic center stars are quite young and do not span a range in metallicity, it is not possible to conclude that the same secondary-like source that drives the chemical evolution in the thin disk is at work.
The observed increase of [F/Fe] with [Fe/H] is not predicted by any of the chemical evolution models discussed and presented in Figure 6, suggesting that quantifying the major source of $^{19}$F at high metallicities remains elusive.

\section{Acknowledgments}

\nolinenumbers
\begin{acknowledgments}
RG acknowledges funding of the PCI fellowship (number 313388/2019-9) at Observatorio Nacional - Rio de Janeiro.
We thank the referee for suggestions that improved the paper.
This work has made use of data from the European Space Agency (ESA) mission {\it Gaia} (\url{https://www.cosmos.esa.int/gaia}), processed by the {\it Gaia} Data Processing and Analysis Consortium (DPAC, \url{https://www.cosmos.esa.int/web/gaia/dpac/consortium}). Funding for the DPAC has been provided by national institutions, in particular the institutions participating in the {\it Gaia} Multilateral Agreement.

{\it Facilities: {Gemini Observatory} and {Keck Observatory}.}

\software{IRAF (Tody \citeyear{tody1986}, Tody \citeyear{tody1993}), Turbospectrum (Alvarez \& Plez \citeyear{alvarez1998}; Plez \citeyear{plez2012}), MOOG (Sneden et al. \citeyear{sneden2012}) and BACCHUS (Masseron, Merle \& Hawkins \citeyear{masseron2016}).}

\end{acknowledgments}

\clearpage

\begin{table}
\centering

\caption{Galactic Center Stars}
\begin{tabular}{llcccccc}
\hhline{========}
          &          &              & Observation & Observation &   &   &   \\ 
Star      & 2MASS ID & Spectrograph & [NIRSpec] & [Phoenix]   & J & H & K${_s}$ \\ \hline
BSD 72    & ...               & NIRSpec/Phoenix & 08/16/2001 & 05/11/2002 & 12.79 & 10.32 & 9.05${^a}$ \\
BSD 124   & 17454240-2859510  & NIRSpec/Phoenix & 08/16/2001 & 05/11/2002 & 14.92 & 10.98 & 8.97     \\
IRS 11    & 17453938-2900148  & Phoenix         & ...        & 05/11/2002 & 15.12 & 10.62 & 8.35     \\
IRS 19    & ...               & NIRSpec         & 08/15/2001 & ...        & 14.81 & 10.83 & 8.22${^a}$ \\
IRS 22    & 17453982-2900539  & NIRSpec         & 05/23/2000 & ...        & 14.20 & 9.65  & 7.39     \\
VR 5-7${^b}$    & 17461658-2849498  & NIRSpec         & 06/13/2001 & ...        & 13.06 & 9.31  & 7.47     \\
\hline
\end{tabular}
\begin{tablenotes}
\item \textbf{Notes}: (a) Soubiran et al. \citeyearpar{soubiran2016}; (b) Member of the Quintuplet cluster
\end{tablenotes}
\label{tab:parameters_center}
\end{table}

\begin{table}
\centering
\caption{Line List}
\begin{tabular}{lclrcc}
\hhline{======}
                 & $\lambda _{air}$ & $\chi$        &             & D$_\circ$ \\ 
Molecule         & (\AA)           &  (eV)          & $\log$ $gf$ & (eV)      \\ \hline
H$^{19}$F(1--0)R9  & 23358.329    & 0.227${^a}$ ${^b}$   & -3.962${^a}$   & 5.869${^c}$ \\
H$^{19}$F(1--0)R11 & 23134.757    & 0.332${^a}$     & -3.942${^a}$   & 5.869${^c}$  \\ 
$^{12}$C$^{16}$O(2--0)R80   & 23109.370       & 1.51${^d}$     & -4.907${^d}$  & 11.092    \\
$^{12}$C$^{16}$O(2--0)R81   & 23122.097       & 1.55${^d}$     & -4.900${^d}$  & 11.092    \\
$^{12}$C$^{16}$O(2--0)R86   & 23192.581       & 1.74${^d}$     & -4.879${^d}$  & 11.092    \\
$^{12}$C$^{16}$O(3--1)R26   & 23341.225       & 0.431${^d}$   & -5.065${^d}$  & 11.092    \\
$^{12}$C$^{16}$O(3--1)R4    & 23367.752       & 0.005${^d}$   & -6.338${^d}$  & 11.092    \\
\hline
\end{tabular}
\begin{tablenotes}
\item \textbf{Notes}: (a) J\"onsson et al. \citeyearpar{jonsson2014a}; (b) Decin \citeyearpar{decin2000}; (c) Sauval \& Tatum \citeyearpar{sauval_tatum1984}; (d) Goorvitch \citeyearpar{goorvitch1994}.
\end{tablenotes}
\label{tab:lines}
\end{table}

\begin{table}
\centering
\scalefont{0.80}
\caption{Stellar Parameters and Abundances}
\begin{tabular}{lcccccccccccc}
\hhline{============}
Star    & T$_{\rm eff}^{a}$ & $\log$g$^{a}$ & $\xi$           & Mass$^{a}$ & R$_g^{a}$ & A(Fe)$^{a}$ &  A(C)  &  $^{16}$O/$^{17}$O  & A(F)$_{R9}$ & A(F)$_{R11} $& $<$A(F)$>$          \\ \hline
BSD 72  & 3880          &  0.80         & 4.15 $\pm$ 0.38 & 5.0        & 1.60      & 7.51        & 8.27 $\pm$ 0.14 & 95$\pm$30 &  4.90 & 4.91 & 4.91 $\pm$ 0.01 \\
BSD 124 & 3735          &  0.40         & 4.84 $\pm$ 0.27 & 6.5        & 1.90      & 7.61        & 8.18 $\pm$ 0.13 & 160$\pm$40 &  4.80 & 4.85 & 4.83 $\pm$ 0.03 \\
IRS 11  & 3625          &  0.30         & 2.30           & 4.5        & 0.62      & 7.53        & 8.40         &    ...  & ...  & 4.38 & 4.38 \\
IRS 19  & 3850          &  0.10         & 3.96 $\pm$ 0.24 & 15.        & 1.00      & 7.63        & 8.16 $\pm$ 0.21 & 390$\pm$60  &  4.85 & 4.95 & 4.90 $\pm$ 0.05 \\
IRS 22  & 3750          &  0.20         & 4.33            & 10.        & 1.00      & 7.57        & 8.09 $\pm$ 0.05 & 170$\pm$30  &  4.74 & 4.75 & 4.75 $\pm$ 0.01 \\ 
VR 5-7  & 3600          & -0.15         & 4.57 $\pm$ 0.07 & 14.        & 31.0      & 7.60        & 7.82 $\pm$ 0.02 & 270$\pm$30  &   4.75            & 4.75 & 4.75 $\pm$ 0.00 \\
\hline
\end{tabular}
\begin{tablenotes}
\item \textbf{Notes}: (a) Cunha et al. \citeyearpar{cunha2007} .
\end{tablenotes}
\label{tab:abundances}
\end{table}

\clearpage

\begin{table}
\centering
\caption{Abundance Sensitivities to Stellar Parameters}
\begin{tabular}{lccccc}
\hhline{======}
Element & $\delta$T$\rm_{eff}$=+100 K & $\delta\log g$=+0.25 & $\delta$[Fe/H]=+0.10 & $\delta\xi$=+0.25 km$\cdot$s$^{-1}$     & $\Delta$A$^{a}$  \\ \hline
F                   & + 0.20 & + 0.02 & + 0.05 & -- 0.02 & $\pm$ 0.21 \\
C (weak CO lines)   & + 0.06 & + 0.13 & + 0.13 & +  0.00 & $\pm$ 0.19 \\
C (strong CO lines) & + 0.06 & + 0.13 & + 0.13 & -- 0.04 & $\pm$ 0.20 \\
\hline
\end{tabular}
\begin{tablenotes}
\item \textbf{Notes}: Baseline model: T$\rm _{eff}$ = 3700 K; log g = 0.50; [Fe/H]=0.00; $\xi$ = 4.50 km/s. (a) Total uncertainty in abundance is defined as $\Delta$A $= [(\delta T)^2 + (\delta \log g)^2 + (\delta [Fe/H])^2 + (\delta \xi)^2]^{1/2}$.
\end{tablenotes}
\label{tab:disturbance}
\end{table}

\clearpage

\end{document}